\documentclass[epj]{svjour}
\usepackage{graphics}
\usepackage{graphicx}
\usepackage{graphicx}%
\usepackage{multirow}%
\usepackage{amsmath,amssymb,amsfonts}%
\usepackage{mathrsfs}%
\usepackage[title]{appendix}%
\usepackage{xcolor}%
\usepackage{textcomp}%
\usepackage{manyfoot}%
\usepackage{booktabs}%
\usepackage{algorithm}%
\usepackage{algorithmicx}%
\usepackage{algpseudocode}%
\usepackage{listings}%
\begin{document}
\title{High-$Q_0$ Treatment of CEBAF 1.5 GHz SRF Cavities}

\author{Pashupati Dhakal
\thanks{dhakal@jlab.org} %
}                     
%
%
\institute{Thomas Jefferson National Accelerator Facility, Newport News, USA}
\date{Received: \today / Revised version: }
%
\abstract{
The Continuous Electron Beam Accelerator Facility (CEBAF) was the first large-scale accelerator to employ superconducting radiofrequency (SRF) cavities for continuous-wave operation. Ongoing research and development efforts continue to focus on increasing the intrinsic quality factor ($Q_0$) of these cavities in order to reduce cryogenic losses while maintaining operational gradients. In this work, we report on the application of high-$Q_0$ surface treatments to single-cell and multicell C100 and C75 style 1.5 GHz niobium cavities used in the CEBAF accelerator. Nitrogen infusion and oxygen alloying via medium-temperature baking were applied under heat-treatment constraints relevant to existing cavity hardware. Both processes yielded substantial improvements in $Q_0$ at moderate accelerating gradients, achieving values of approximately 2 $\times$ 10$^{10}$ at 2.07 K and 20 MV/m. The effectiveness of nitrogen infusion at reduced annealing temperatures and the successful extension of oxygen alloying to multicell cavities are demonstrated. These results establish viable pathways for implementing high-$Q_0$ treatments in CEBAF-compatible cavities and support future integration into cryomodules for reduced operational cryogenic load
\PACS{
      {PACS-key}{SRF Cavity}   \and
      {PACS-key}{Particle Accelerator} \and
       {PACS-key} {High Q$_0$}
     } 
} 
\maketitle
\section{Introduction}
\label{intro}
The Continuous Electron Beam Accelerator Facility (CEBAF) at Jefferson Lab is a pioneering accelerator that employs large-scale superconducting radiofrequency (SRF) cavities to deliver high-duty-factor electron beams for nuclear physics experiments \cite{leemann}. As of August 2025, CEBAF operates at 2.07 K with a total of 52 cryomodules that are equally distributed in two linacs, each housing eight multicell SRF cavities operating at 1.5 GHz, providing a combined beam energy of approximately 1060 MeV per linac.

Over the past decade, substantial progress has been made in SRF cavity surface processing to enhance the intrinsic quality factor ($Q_0$) at moderate accelerating gradients ($\sim$20–25 MV/m). Improvements in $Q_0$ directly reduce RF dissipation and cryogenic load, which is critical for continuous-wave accelerators such as CEBAF. Early upgrades replaced the original buffer chemical polished (BCP) C20 cryomodules with C50 units that adopted electropolishing (EP) as the final surface treatment, resulting in higher achievable gradients. The 12 GeV CEBAF upgrade further introduced low-loss seven-cell cavities housed in individual helium vessels, forming the C100 cryomodules \cite{12Gev} . These cavities were electropolished and subsequently treated with a low-temperature bake (LTB) at 120 $^\circ$C for 48 hours \cite{andrew}, which effectively mitigated the high-field Q-slope but yielded only moderate improvements in $Q_0$ due to competing changes in the BCS and residual resistance components \cite{bashuIEEE}.

The development of nitrogen alloying in 2013 marked a major advance in SRF performance \cite{anna2013}. This process involves high-temperature furnace treatment in a nitrogen atmosphere, followed by controlled material removal via electropolishing, and was later industrialized for the LCLS-II project \cite{dan}. Nitrogen alloying enabled production-average $Q_0$ values exceeding 3 $\times$ 10$^{10}$ at 16 MV/m, with optimal performance strongly dependent on efficient magnetic flux expulsion during cooldown \cite{sam19}. Achieving reliable flux expulsion typically requires high-temperature annealing at temperatures approaching 975 $^\circ$C, which is incompatible with certain cavity assemblies.

An alternative approach, nitrogen infusion, introduces nitrogen at lower temperatures during furnace cooldown and eliminates the need for post-treatment electropolishing. This method has demonstrated exceptionally high $Q_0$ values with increasing accelerating gradient and compatibility with high-gradient operation in single-cell cavities \cite{anna,dhakalsrf19,dhakalreview,dhakalinfusion}. However, reproducibility of nitrogen infusion in multicell cavities remains an active area of investigation.

More recently, oxygen alloying through medium temperature baking (mid-T bake) has emerged as a promising high-$Q_0$ treatment \cite{sam,Fhe,eric,dhakalipac24}. This process exploits the diffusion of oxygen from the native oxide into the near-surface region of niobium during moderate heat treatments (200–400 $^\circ$C), modifying the oxide and suboxide states without introducing external dopants \cite{Prudnikava}. Oxygen alloying has been successfully demonstrated in multicell cavities and offers reduced sensitivity to trapped magnetic flux, while avoiding high-temperature furnace cycles and chemical post-processing \cite{pan}.

Each of these surface treatments presents distinct advantages and limitations in terms of achievable $Q_0$, operational gradient, reproducibility, and compatibility with existing cavity hardware. In particular, cavities currently deployed in CEBAF C100 cryomodules contain stainless-steel components brazed to the beam tubes, restricting allowable heat-treatment temperatures to  $\le$ 600 $^\circ$C and limiting the applicability of high-temperature nitrogen alloying. In contrast, refurbished C20 cryomodules and newly fabricated C75 cavities \cite{c75} constructed entirely of niobium and processed prior to helium vessel installation—are suitable candidates for advanced high-$Q_0$ treatments.

Motivated by these considerations, this work investigates the application of nitrogen infusion and oxygen alloying to both single-cell and multicell cavities of C100 and C75 geometries relevant to CEBAF. The performance gains, flux trapping sensitivity, and operational implications of these treatments are systematically evaluated with the goal of identifying viable high-$Q_0$ processing routes compatible with CEBAF refurbishment and upgrade programs.

\section{Surface Preparation}
Single cell and multicell cavities used in CEBAF accelerators, made from high purity fine and large grain Nb were processed with 25 $\mu$m electropolishing for baseline measurements. Both large grain and find grain cavities showed similar performance when treated with standard surface polishing \cite{kneisel}.  These cavities were previously heat treated multiple times during ongoing R\&D work. After the baseline measurement, the cavities were high pressure rinsed and clean Nb caps were installed in clean room and transported to the furnace room sealed with clean-room plastic bag. 
Nitrogen infusion was performed by annealing the cavities at either 600 or 800 $^\circ$C/3h followed by injection of nitrogen gas to $\sim$25 mTorr at $\sim$300 $^\circ$C during the cooldown. The partial pressure is maintained while the furnace cooldown to $\sim$ 165 $^\circ$C, measured at cavity surface. The temperature and pressure was held at 165 $^\circ$C for additional 48 hours \cite{dhakalsrf19,dhakalinfusion}. The cavities surface was reset by 25 $\mu$m electropolishing prior to all repeated infusion runs.  Table \ref{tab:cavity} lists the cavity ID and parameters in our current study.

\begin{table*} [htb]
   \centering
   \caption{Cavity ID, geometry type, material, cell count, and geometry dependent cavity parameters: $B_p/E_{acc}$ and geometric factor (G)}
   \begin{tabular}{@{}llllll@{}}
     \toprule
       \textbf{Cavity ID} & \textbf{Cavity Shape} &\textbf{Material Type}  & \textbf{\# of Cells } & \textbf{B$_p$/E$_{acc}$ (mT/(MV/m)} & \textbf{G}\\
       \midrule
           RDL-02         & C100  & Fine grain  &  1   & 4.18 & 277.2 \\
            C75-SC1         & C75  &Large grain  &  1  & 4.18 & 277.2 \\
            C75-RI-37        & C75  & Large grain   &  5   & 4.22 & 275.4 \\
              C75-RI-47         & C75  &Large grain    &  5  & 4.22 & 275.4 \\ 
               C100-01         & C100  & Fine grain   & 7   & 3.89 & 281.0 \\
     \bottomrule
   \end{tabular}
      \label{tab:cavity}
        
   \end{table*}
  
The cavity's surface was again reset by 25 $\mu$m electropolishing for baseline measurements for O-alloying. The cavity preparation for O-alloying also involves the high pressure rinse, clean Nb foils installation and furnace treatment. The cavities were heat treated at $\sim$ 350 $^\circ$C for 3 hours. The temperature and duration were chosen based on the previous studies which gives high $Q_0$ and high $E_{acc}$ with lower flux trapping sensitivity \cite{dhakalipac24}.
All cavities were tested twice: first with minimal residual magnetic field in the Dewar ($<$ 2 mG) and second with the 20 mG in full flux trapping condition. The full flux trapping condition was obtained by applying 20 mG field in the vertical Dewar with magnetic field compensation coil along the cavity beam-tube axis and slow cooldown through the superconducting transition temperature ($T_c$). The increase in surface resistance as a result of residual trapped flux is referred as flux trapping sensitivity (S).

\section{Results}
The summary of the rf results were presented in Table \ref{tab:rftest}.
\begin{table*} [htb]
   \centering
   \caption{Summary of RF test results on single cell and multi-cell CEBAF cavities at 2.07 K.}
   \begin{tabular}{@{}lllll@{}}
     \toprule
       \textbf{Cavity ID} & \textbf{Treatment} &\textbf{E$_{max}$( MV/m)}  & \textbf{$Q_0$ at E$_{max}$  } & \textbf{Q$_{0, max}$}\\
       \midrule
           RDL-02         & EP  & 32  &  3.8$\times10^{9} $  &  1.2$\times10^{10} $   \\
           RDL-02         & EP+800 $^\circ$C/3h+165 $^\circ$C/48h with N$2$  & 22  &  2.0$\times10^{10} $  &  2.0$\times10^{10} $   \\
           RDL-02         & EP+600 $^\circ$C/3h+165 $^\circ$C/48h with N$2$  & 25  &  1.85$\times10^{10} $  &  2.0$\times10^{10} $   \\
           RDL-02         & EP+350 $^\circ$C/3h  & 32  &  3.1$\times10^{10} $  &  2.4$\times10^{10} $   \\
            C75-SC1         & EP+600 $^\circ$C/3h+165 $^\circ$C/48h with N$2$  &20  &  1.4$\times10^{10}$  & 1.6$\times10^{10}$  \\
            C75-SC1         & EP+350 $^\circ$C/3h &16 &  1.87$\times10^{10}$  & 2.0$\times10^{10}$  \\
            
            C75-RI-37        & EP+350 $^\circ$C/3h & 20   &  1.9$\times10^{10}$  & 2.0$\times10^{10}$ \\
              C75-RI-47         & EP+350 $^\circ$C/3h  &26    &  1.7$\times10^{10}$  & 1.9$\times10^{10}$  \\ 
               C100-01         & EP  & 25  & 0.9$\times10^{10}$   & 1.3$\times10^{10}$ \\
                C100-01         & EP+600 $^\circ$C/3h+165 $^\circ$C/48h with N$2$  & 25  & 1.86$\times10^{10}$   & 1.9$\times10^{10}$ \\
                 C100-01         & EP+350 $^\circ$C/3h  & 22  & 2.0$\times10^{10}$   & 2.3$\times10^{10}$ \\
     \bottomrule
   \end{tabular}
      \label{tab:rftest}
        
   \end{table*}

\subsection{Nitrogen Infusion}
\begin{figure}
    \centering
    \includegraphics[width=0.95\linewidth]{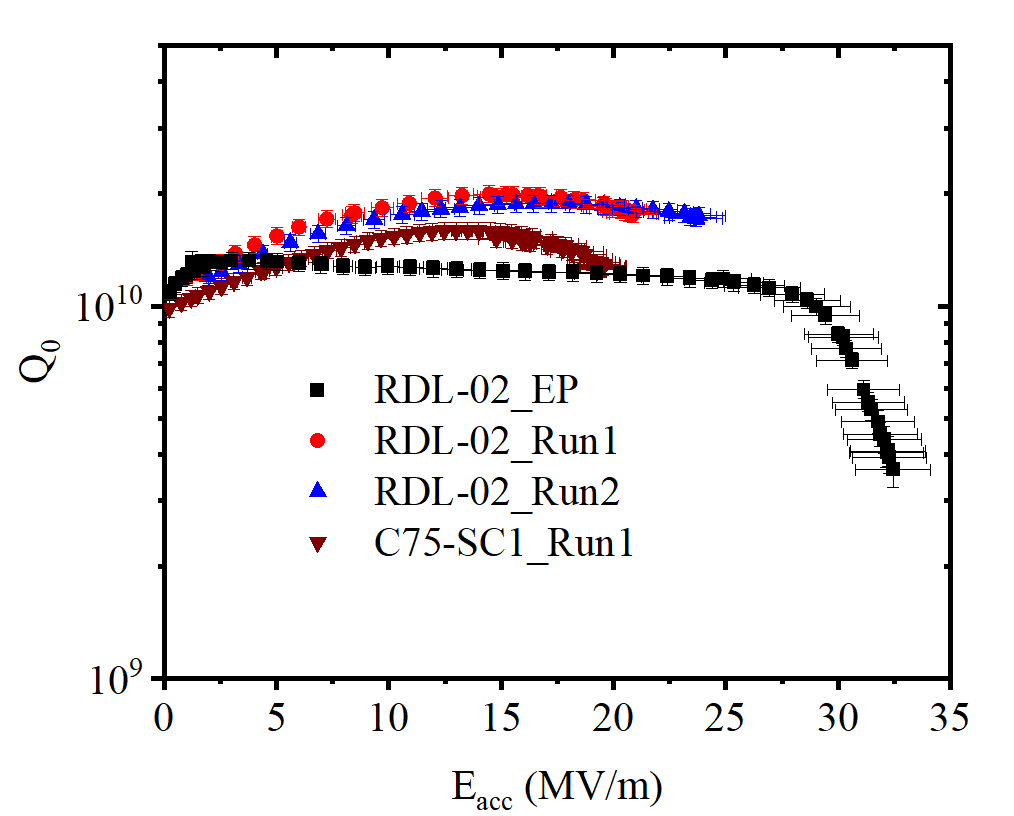}
    \caption{$Q_0(E_{acc})$  of two single cell cavities at 2.07 K after nitrogen infusion.}
    \label{fig:single-cell-N}
\end{figure}

Figure \ref{fig:single-cell-N} shows the $Q_0(E_{acc}$) of single cell cavities at a 2.07 K and residual magnetic field $<$ 2 mG in the Dewar. A reference baseline test for cavity RDL-02 with EP was limited by a high-field Q-slope at approximately 32 MV/m. The first infusion run involved an 800 $^\circ $C/3h furnace treatment followed by a 165 $^\circ$C/48h nitrogen infusion. After the first nitrogen infusion, the cavity exhibited a higher Q$_0$ of approximately 2 $\times$ 10$^{10}$ at 20 MV/m, before it quenched at 22 MV/m. These results confirm the reproducibility of previous studies \cite{dhakalinfusion}. A second infusion run was carried out with a modified degassing process, lowering the furnace temperature to 600 $^\circ $C/3h, followed by the same 165 $^\circ $C/48 h nitrogen infusion. In this case, the cavity reached an E$_{acc}$ of  25 MV/m with a Q$_0$ of 2$\times$10$^{10}$ at 20 MV/m. The infusion run with 600 $^\circ $C/3h heat treatment was also done on C75-SC1 which also showed the increase in $Q_0$ as accelerating gradient increases, however the overall quality factor is lower than RDL-02.

\begin{figure}
    \centering
    \includegraphics[width=1.1\linewidth]{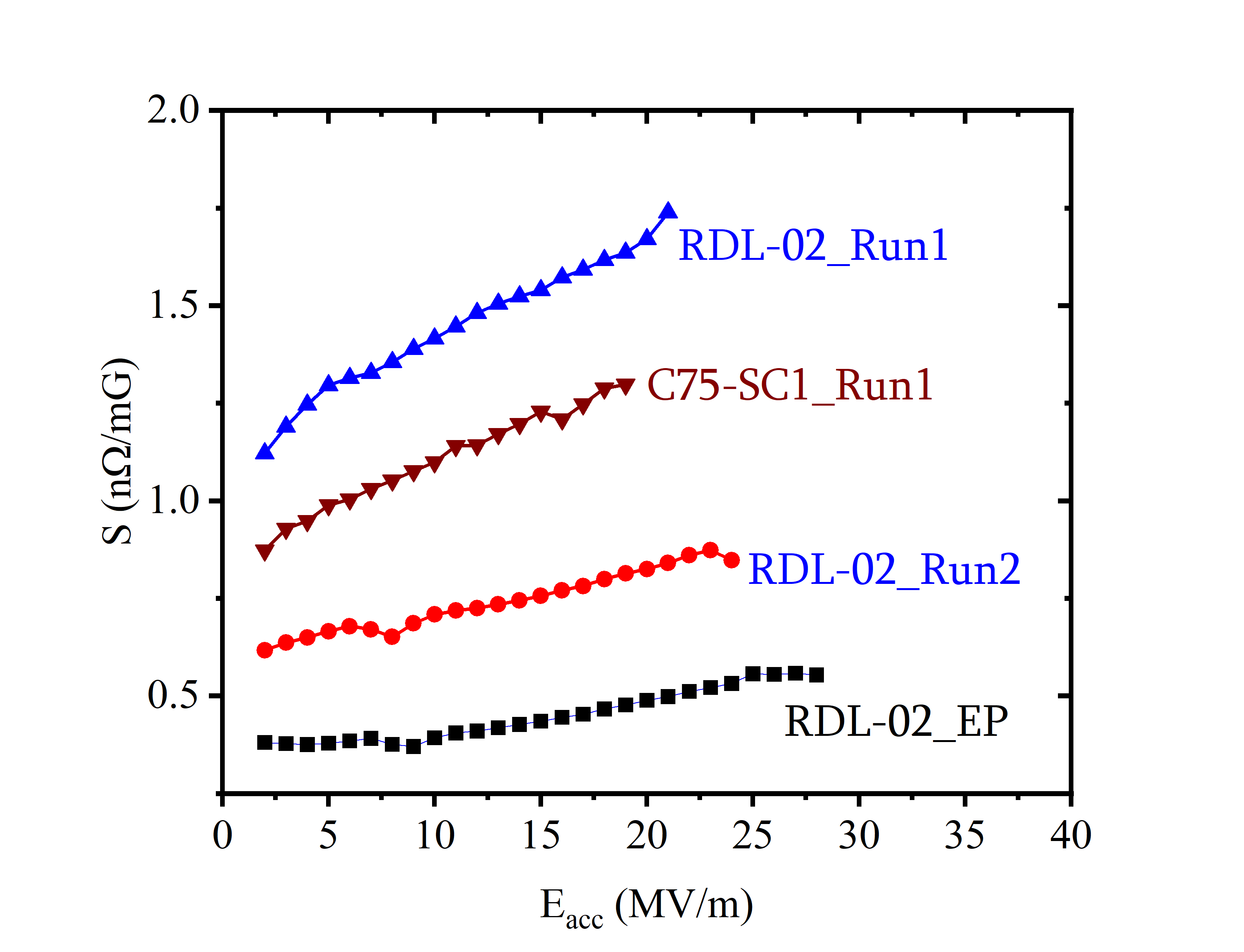}
    \caption{Flux trapping sensitivity of single cell cavities after nitrogen infusion with different annealing temperature as described in text.}
    \label{fig:single-cell-N-S}
\end{figure}

Figure \ref{fig:single-cell-N-S} shows the flux trapping sensitivity as a function of accelerating gradient. The flux trapping sensitivity as a function of accelerating gradient showed a substantial variation among cavities and treatment. 

\begin{figure}
    \centering
    \includegraphics[width=0.95\linewidth]{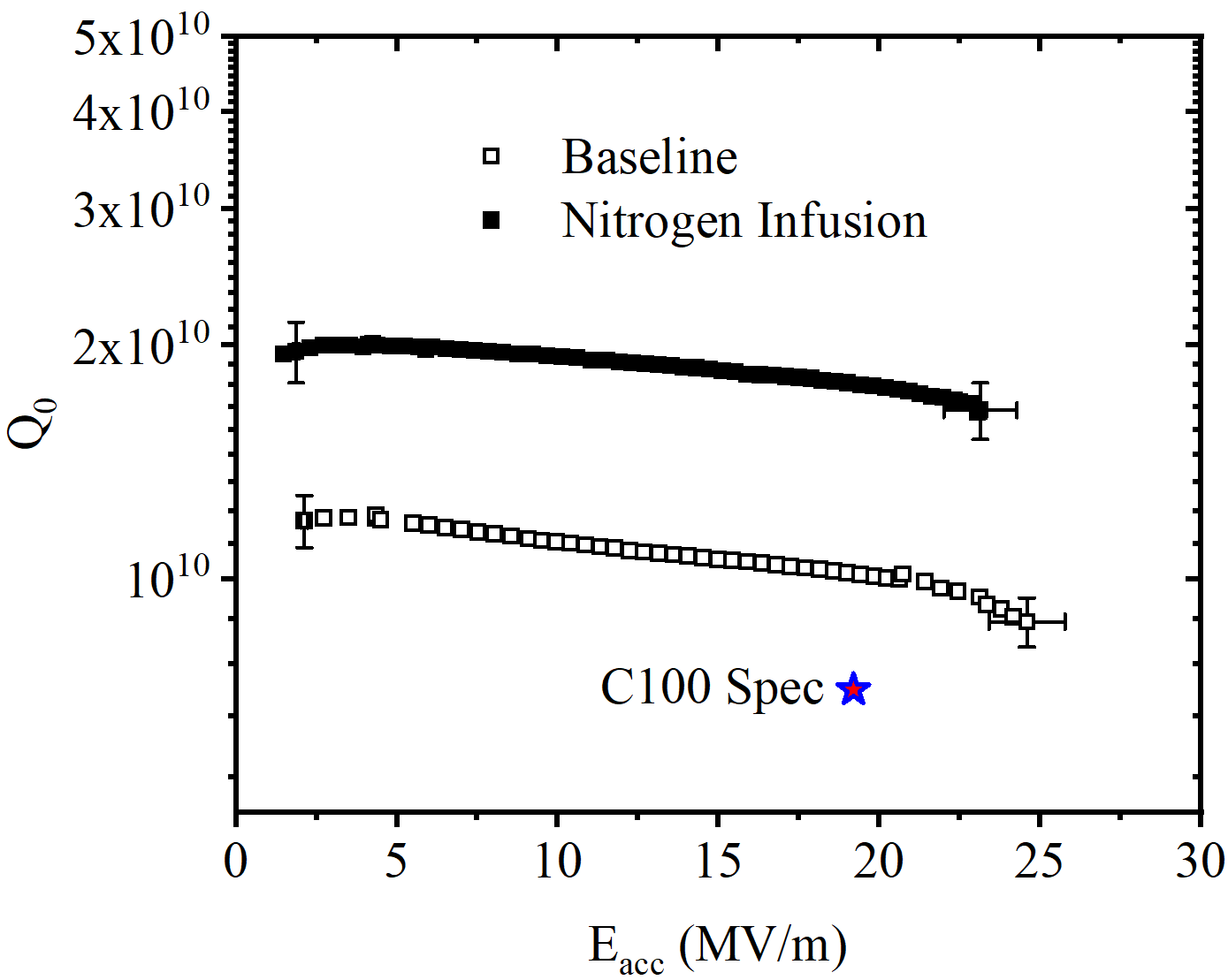}
    \caption{$Q_0(E_{acc})$ of 7-cell cavities after nitrogen infusion with 600 $^\circ $C/3h furnace treatment and 165 $^\circ $C/48 h process at 2.07 K. The star symbol represents the C100 specification.}
    \label{fig:C100-7cell}
\end{figure}

Following the successful infusion process on single-cell cavities, the same 600 $^\circ $C/3h furnace treatment and 165 $^\circ $C/48 h infusion run was applied to a 7-cell C100-shaped cavity. Figure \ref{fig:C100-7cell} shows the Q$_0$(E$_{acc}$) at 2.07 K after the baseline EP treatment and the nitrogen infusion. The 7-cell cavity achieved an E$_{acc}$ $\sim$ 24 MV/m, with a Q$_0$ of 1.86 $\times$ 10$^{10}$ at 20 MV/m, which represents an 80\% increase in $Q_0$ compared to the baseline test. 

\subsection{Oxygen Alloying (Mid-T bake)}
\begin{figure}
    \centering
    \includegraphics[width=0.95\linewidth]{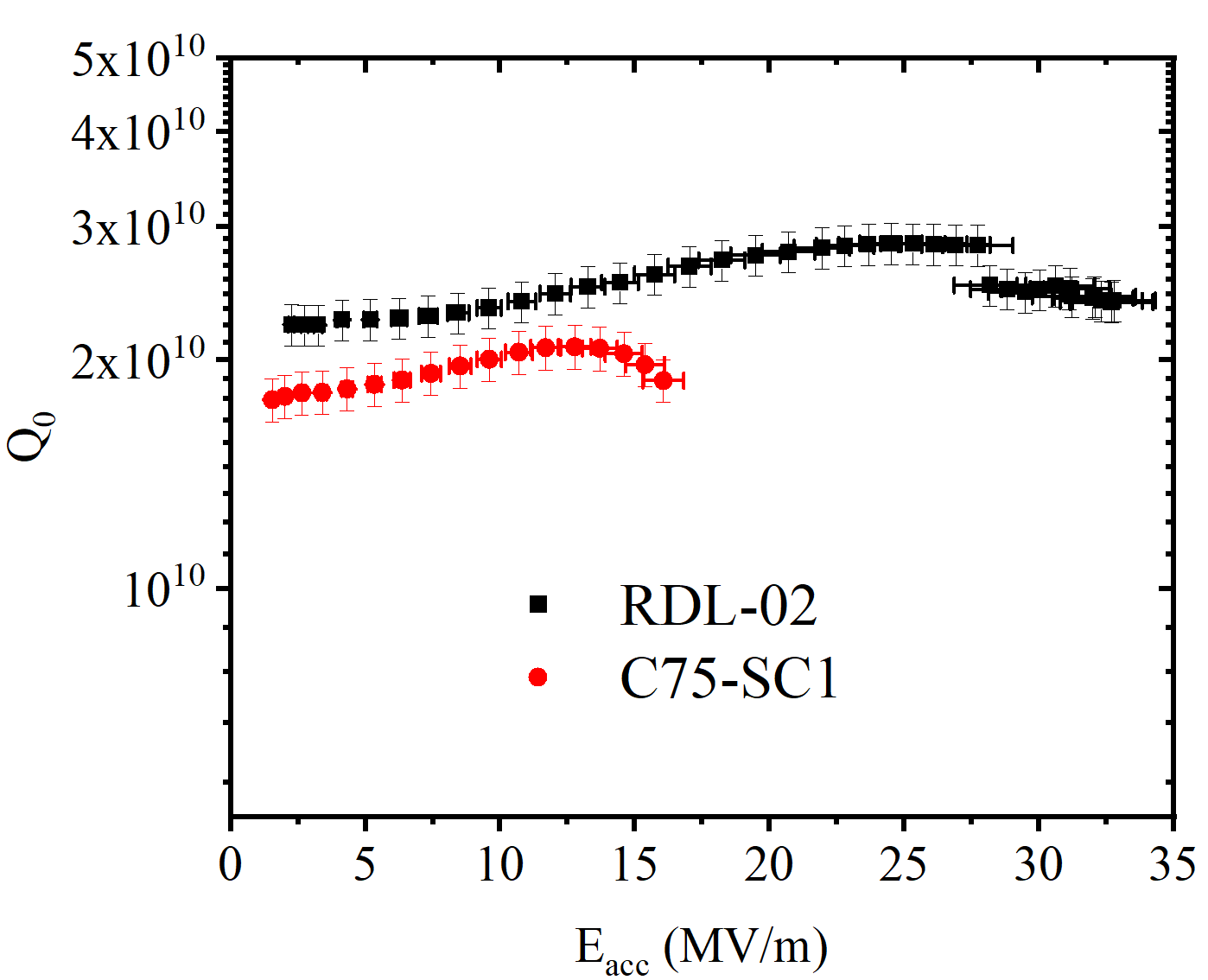}
    \caption{$ Q_0(E_{acc})$ of two single cell cavities after oxygen alloying at 350 $^\circ$C for 3 hours measured at 2.07 K.}
    \label{fig:midT_single}
\end{figure}

Single cell cavities were electropolished $\sim$ 25 $\mu$m in order to surface reset and treated with medium temperature bake. The heat treatment temperature of 350 $^\circ$C for 3 hours was chosen based on the previous studies in 1.3 GHz cavities, which provides the high $Q_0$ and $E_{acc}$ along with the lower flux trapping sensitivity \cite{dhakalipac24}. As shown in Fig. \ref{fig:midT_single}, both cavities showed Q-rise with cavity RDL-02 reaching to $E_{acc} = $ 32 MV/m with $Q_0=$ 2.4$\times$10$^{10}$. The decrease in $Q_0$ was observed at 28 MV/m due to multipacting, which is eventually processed with high rf power. Cavity C75-SC1 was limited at $E_{acc}$=  16 MV/m with $Q_0=$ 1.87$\times$10$^{10}$. 

\begin{figure}
    \centering
    \includegraphics[width=0.9\linewidth]{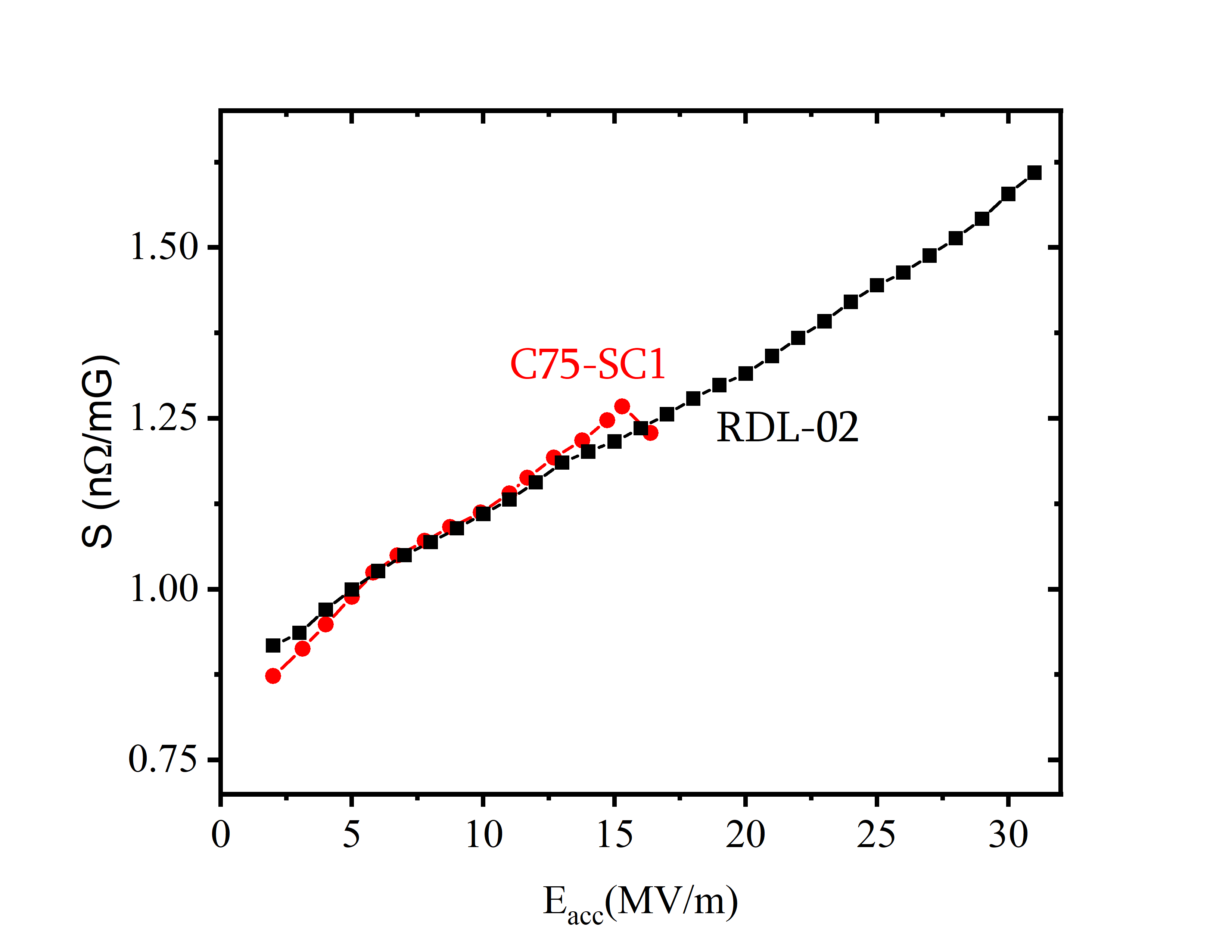}
    \caption{Flux trapping sensitivity of single cell cavities after oxygen alloying at 350 $^\circ$C for 3 hours measured at 2.07 K}
    \label{fig:single-cell-m-S}
\end{figure}

Figure \ref{fig:single-cell-m-S} shows the flux trapping sensitivity of single cell cavities at 2.07 K after the mid-T bake. Both cavities shows the similar flux trapping sensitivity as a function of accelerating gradient. 

\begin{figure}
    \centering
    \includegraphics[width=0.95\linewidth]{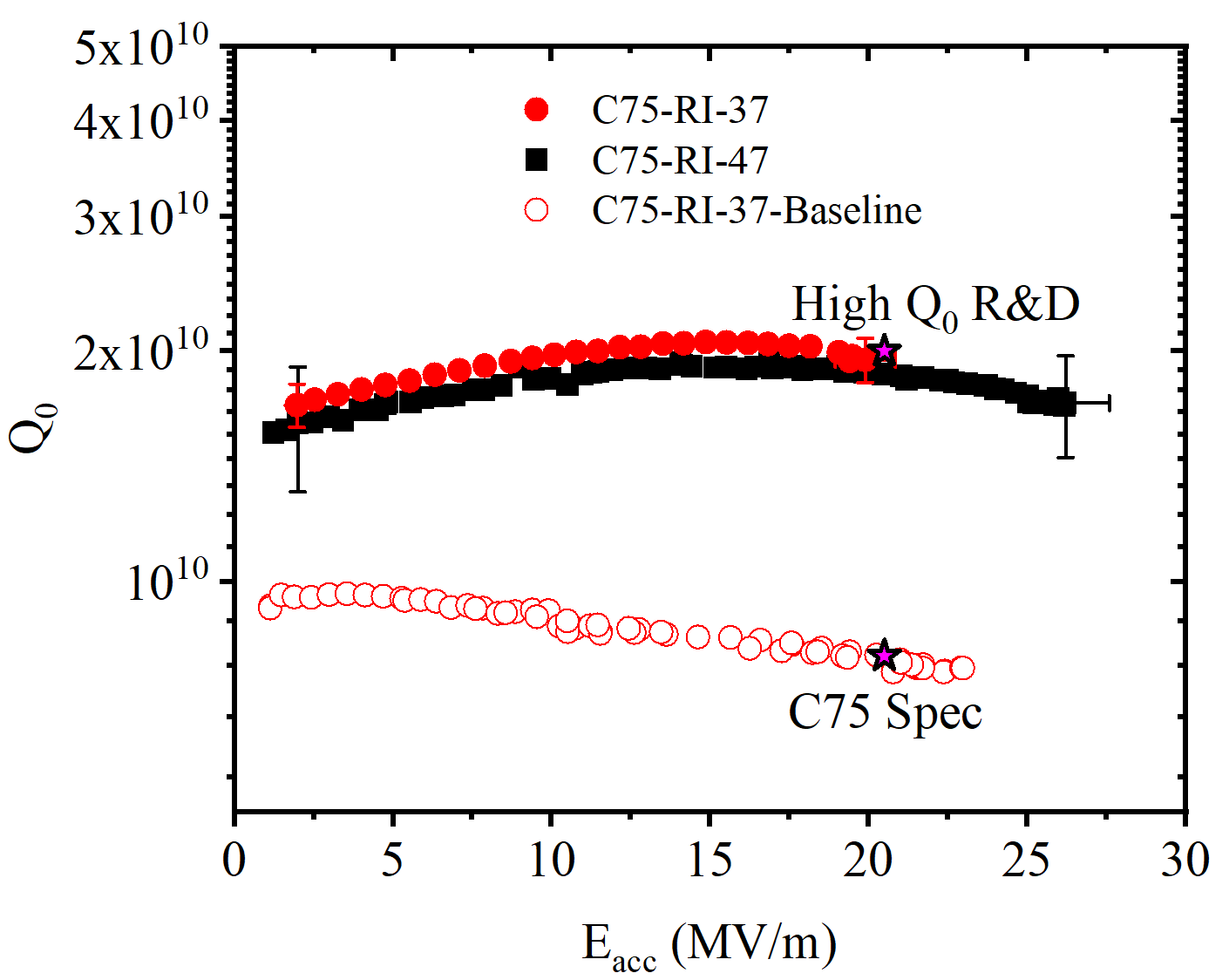}
    \caption{$Q_0(E_{acc})$ of 5-cell cavities after oxygen alloying at 350 $^\circ$C for 3 hours measured at 2.07 K. The star symbol represents the C75 and current R\&D specification.}
    \label{fig:C75-5cell}
\end{figure}

\begin{figure}
    \centering
    \includegraphics[width=0.95\linewidth]{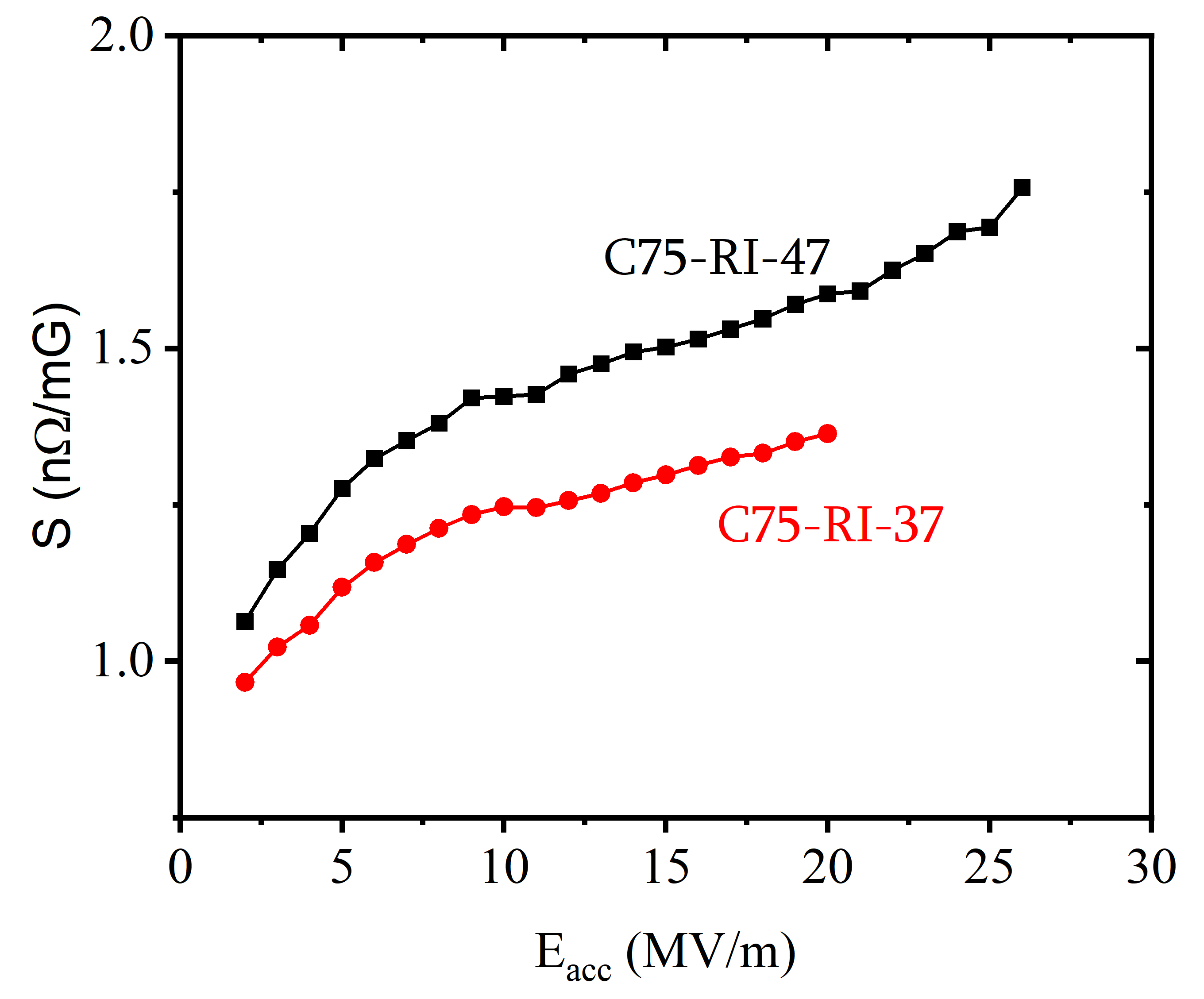}
    \caption{Flux trapping sensitivity of 5-cell cavities after oxygen alloying at 350 $^\circ$C for 3 hours measured at 2.07 K.}
    \label{fig:5-cell-s}
\end{figure}

After the successful demonstration of high $Q_0$ in previous 1.3 GHz single cell cavities \cite{dhakalipac24} and current 1.5 GHz cavities, the same baking step was applied to 5 cell cavities. It is to be noted that the 5-cell cavities were fabricated using large grain niobium. Figure \ref{fig:C75-5cell} shows the $Q_0(E_{acc})$ of two 5 cell cavities measured at 2.07 K. Both cavities showed the Q-rise feature. Cavity C75-RI-37 reached  $E_{acc}$ of 20 MV/m with $Q_0$ = 1.9$\times$10$^{10}$, whereas cavity C75-RI-47 reached $E_{acc}$ of 26 MV/m with $Q_0$ = 1.7$\times$10$^{10}$. 
Figure \ref{fig:5-cell-s} shows the flux trapping sensitivity of 5-cell C75 cavities after mid-T bake at 2.07 K. A small variation in flux trapping sensitivity was observed between two cavities treated with the same process. These two cavities are currently being prepared to integrate in full cryomodules.

\begin{figure}
    \centering
    \includegraphics[width=0.95\linewidth]{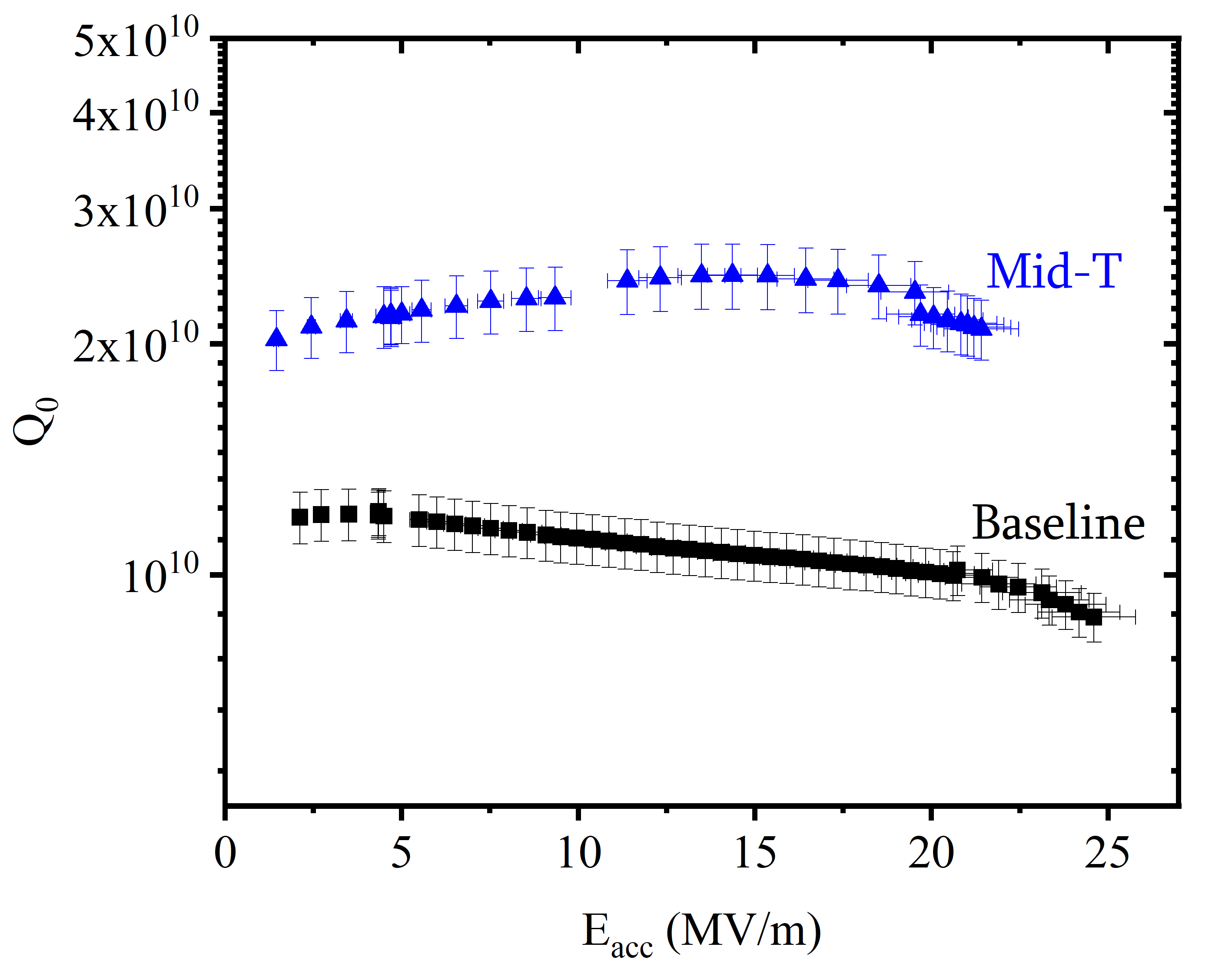}
    \caption{$Q_0(E_{acc})$ of 7-cell cavity after oxygen alloying at 350 $^\circ$C for 3 hours measured at 2.07 K.}
    \label{fig:C100-midT}
\end{figure}

The same mid-T bake process was also applied to C100-01, 7-cell cavity after the surface is reset by $\sim 10~  \mu$m EP. The cavity reached $E_{acc}$ = 21.4 MV/m with $Q_0$ = 2.1$\times$10$^{10}$ as shown in Fig. \ref{fig:C100-midT}. The cavity encountered multipacting $\sim$ 19.5 MV/m but processed quickly, with slight decrease in $Q_0$. Overall more than 100 \% increase in $Q_0$ was observed compared to the baseline test.

\begin{figure}
    \centering
    \includegraphics[width=0.95\linewidth]{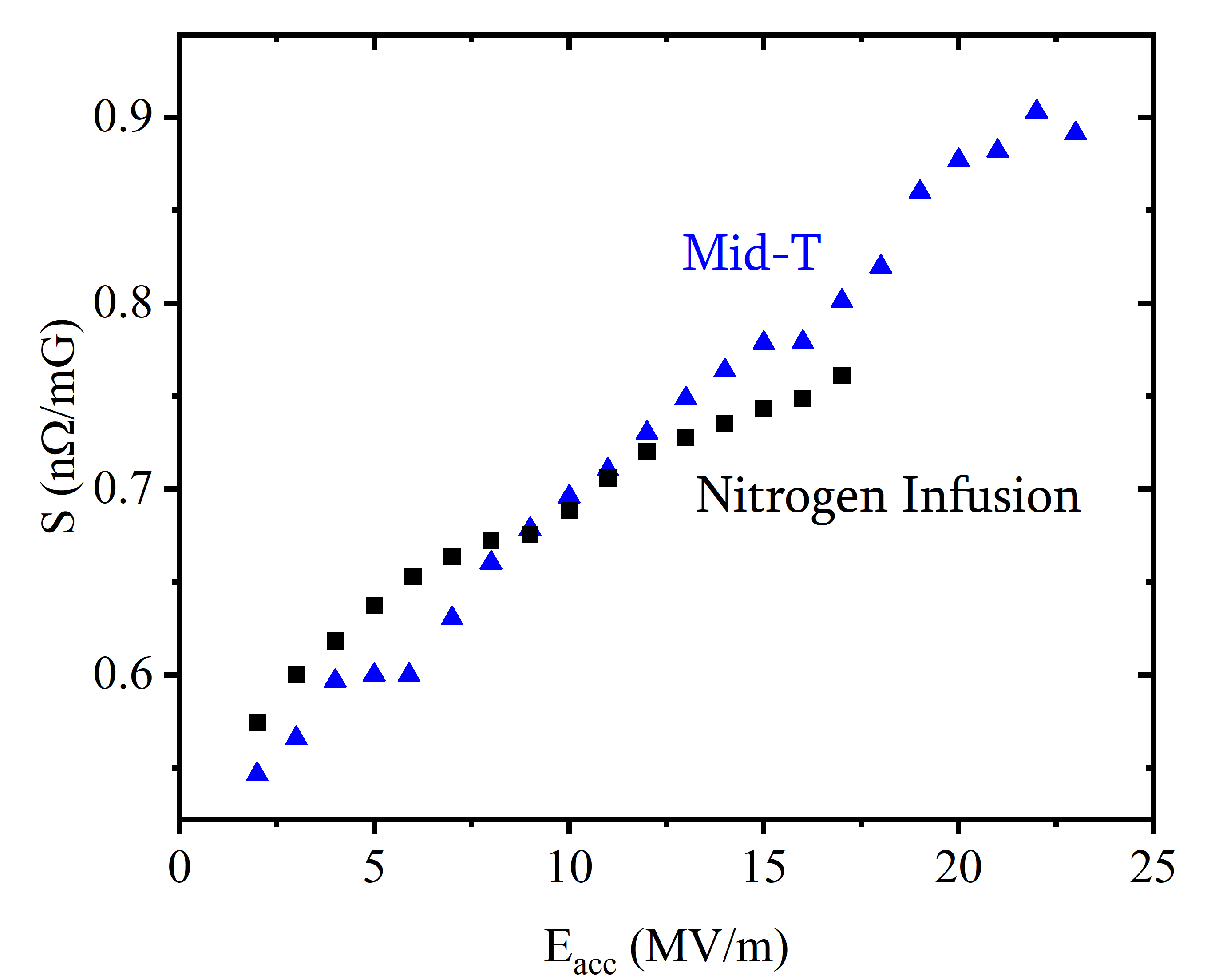}
    \caption{Flux trapping sensitivity of 7-cell cavity after nitrogen infusion at 165 $^\circ$C/48 h and oxygen alloying at 350 $^\circ$C for 3 hours and measured at 2.07 K.}
    \label{fig:C100-S}
\end{figure}

Figure \ref{fig:C100-S} shows the flux trapping sensitivity of 7 cell C100 shaped cavities after the nitrogen infusion and mid-T bake. The flux trapping sensitivity is very similar after the high-$Q_0$ treatment.

\section{Discussion}
While the effects of nitrogen infusion and oxygen alloying are well established for 1.3 GHz cavities, our results show that these treatments similarly enhance the performance of 1.5 GHz cavities important to CEBAF. Although the present measurements were performed at 1.5 GHz and 2.07 K, the achieved $Q_0$ values and flux sensitivities compare favorably with state-of-the-art results reported at 1.3 GHz and 2.0 K. Accounting for the expected frequency scaling of the BCS surface resistance ($\propto f^2$) and at T = 2.07 K, the $Q_0$ $\sim 2\times$10$^{10}$ achieved here at 20 MV/m is fully consistent with nitrogen alloyed and mid-T baked cavities reporting $Q_0\sim2\times$10$^{10}$at 1.3 GHz and 2.0 K.

The study showed the increased flux trapping sensitivity across nitrogen-infused and oxygen-alloyed 1.5 GHz cavities compared to baseline.  Baseline electropolished cavities showed flux trapping sensitivity of $\leq$ 0.5 n$\Omega$/mG and showed moderate field dependence, whereas both nitrogen and oxygen alloyed cavities showed strong field dependence. The increase in sensitivity translates directly to the reduced mean free path as well as the increase in pinning centers within rf penetration depth \cite{dhakalprab}. Using the measured flux sensitivities, it is possible to estimate the operational impact of residual magnetic fields typical of CEBAF cryomodules. For a representative sensitivity of $\sim$1–1.5 n$\Omega$/mG, a residual field of 10 mG would contribute an additional residual resistance of $\sim$10-15 n$\Omega$. At 2.07 K and 1.5 GHz, this corresponds to $\sim$ 40-50\% $Q_0$ degradation. In this regime, the intrinsic Q-rise observed in vertical tests should largely translate to installed cryomodule performance, provided reasonable flux expulsion is achieved. Nevertheless, even under these less favorable conditions, both nitrogen-infused and oxygen-alloyed cavities are expected to outperform baseline EP cavities in cryomodules, offering tangible cryogenic savings. These estimates are consistent with operational experience from LCLS-II \cite{dan} and recent mid-T bake demonstrations \cite{dhakalipac24}, where the highest gains from high-$Q_0$ treatments were realized only when residual fields were controlled below the 5 mG level and strong flux expulsion was achieved during cooldown. 

\section{Summary}
Nitrogen infusion and oxygen alloying via medium temperature baking were systematically applied to several single-cell and multicell 1.5 GHz SRF cavities relevant to the CEBAF accelerator. The results demonstrate that nitrogen infusion remains an effective approach for enhancing $Q_0$ even when the initial annealing temperature is reduced from 800 $^\circ$C to 600 $^\circ$C , making it compatible with cavities that have heat-treatment constraints. However, noticeable cavity-to-cavity variations in flux trapping sensitivity were observed, suggesting sensitivity to extrinsic factors such as furnace conditions, handling during niobium cap installation, and residual contamination.

Oxygen alloying (mid-T bake) was shown to be equally effective in achieving high $Q_0$ across both single-cell and multicell cavities, including those fabricated from large-grain niobium. This process consistently produced Q-rise behavior and provided $Q_0$ values comparable to nitrogen infusion, while avoiding high-temperature furnace treatments and post-processing electropolishing. For multicell C75 and C100 cavities, oxygen alloying resulted in $Q_0$ improvements exceeding 100\% relative to baseline electropolished performance at operational gradients.

Overall, these results indicate that, under realistic residual field conditions of 5–10 mG, both treatments can deliver substantial cryogenic load reductions in CEBAF cryomodules. The study further suggest that achieving and maintaining low ambient magnetic fields may yield performance gains comparable to, or greater than, incremental refinements in surface processing alone. Future efforts will focus on integrating these treated cavities into full cryomodules to evaluate operational stability, cryogenic load reduction, and reproducibility in accelerator environments.

\section{ACKNOWLEDGMENTS}
We would like to acknowledge Jefferson Lab technical staff members for the cavity fabrication, processing, and cryogenic support during RF test. The work is supported by the U.S. Department of Energy, Office of Science, Office of Nuclear Physics under contract DE-AC05-06OR23177.

%
%
\bibliographystyle{naturemag}
\bibliography{references}

\end{document}